\documentclass[a4paper]{jpconf}
\usepackage{graphicx}
\usepackage{slashed}

\begin{document}

\title{Neutrino masses beyond the minimal seesaw}

\author{Ricardo Cepedello$^\dagger$, Renato Fonseca$^*$,
  Martin Hirsch$^{\dagger,}$\footnote{Speaker}}

\address{$^\dagger$ AHEP Group, Instituto de F\'{i}sica Corpuscular
        - CSIC/Universitat de Val\`{e}ncia, \\
        Calle Catedr\'{a}tico Jos\'{e} Beltr\'{a}n, 2
        E-46980 Paterna, Spain}

\address{$^*$ Institute of Particle and Nuclear Physics, Faculty of
  Mathematics and Physics, Charles University, V Holešovi\v{c}kách 2, 18000
  Prague 8, Czech Republic}

\ead{ricepe@ific.uv.es, fonseca@ipnp.mff.cuni.cz, mahirsch@ific.uv.es}

\begin{abstract}
  The simplest possibility to generate small Majorana neutrino masses
  is the seesaw mechanism. However, the smallness of the observed
  neutrino masses can also be understood, if neutrino masses are
  generated by higher-dimensional operators and/or at higher loop
  level. In this talk recent work on systematic classifications of
  higher-dimensional and radiative neutrino mass models is
  summarized. Two particular classes of {\em special genuine} loop
  diagrams, i.e. diagrams which can lead to genuine neutrino mass
  models only under some specific, well-defined conditions, are also
  discussed.

\end{abstract}

\section{Introduction}

At low energy, all Majorana neutrino mass models can be described
effectively by the Weinberg operator \cite{Weinberg:1979sa} or its
higher-dimensional variants ${\cal O}_{d=5+2n}= {\cal O}_W
(HH^{\dagger})^n$. Opening the Weinberg operator at tree-level leads
to three well-known realizations of the seesaw \cite{Ma:1998dn}.
However, many models beyond these minimal seesaw mechanisms have been
proposed in the literature, with different suppression mechanisms to
explain the observed smallness of neutrino masses. In full generality
we can write~\cite{Bonnet:2012kz}:
\begin{equation}
  \label{eq:mnugen}
  m_{\nu} \propto 
  \epsilon 
  \cdot
  \left( \frac{1}{16 \pi^{2}} \right)^{n}
  \cdot 
  \left(\frac{v}{\Lambda}\right)^{d-5}
  \cdot
  \frac{v^2}{\Lambda} .
\end{equation}
Here, $v$ is the vacuum expectation value of the Higgs, $d$
stands for the dimension of the operator, $n$ is the number of loops
at which neutrino masses are generated and $\epsilon$ expresses
symbolically possible additional suppression factors.
Finally, in addition, small couplings not shown explicitly in
Eq. (\ref{eq:mnugen}) could lead to smaller neutrino masses than
naively expected.

There have been several attempts to systematically analyze the
generation of Majorana neutrino masses. The authors of
\cite{Bonnet:2012kz} have found all 1-loop topologies and diagrams at
$d=5$ level, the corresponding ($d=5$) 2-loop study was published in
\cite{Sierra:2014rxa}. Neutrino masses have also been studied at
$d=7$.  A complete tree-level analysis for $d=7$ was given in
\cite{Bonnet:2009ej}, the extension to $d=7$ 1-loop can be found in
\cite{Cepedello:2017eqf}. In this talk, we will mainly discuss two
recent papers: \cite{Anamiati:2018cuq} presented a systematic analysis
of $d=9-13$ tree-level neutrino mass models, while
\cite{Cepedello:2018rfh} gives a complete description of 3-loop
realizations of the Weinberg operator. We also comment on special
genuine diagrams, not discussed in detail in \cite{Sierra:2014rxa}.

\section{High-dimensional tree-level models}

Finding all possible topologies and diagrams for generating neutrino
masses at tree-level and a given $d$ is a straight-forward but tedious
task, because the number of possible topologies grows quickly
with $d$ \cite{Anamiati:2018cuq}, see table \ref{tab:topoa}. Here,
topology is used for Feynman diagrams in which fermions and scalars
are not identified, while diagrams are descended from topologies by
distinguishing the scalar or fermion nature of the lines.

Most of the diagrams at larger dimensions, however, do not lead to
interesting models. Already at $d=7$ one can find examples of diagrams for the
inverse seesaw \cite{Mohapatra:1986bd} or linear seesaw
\cite{Akhmedov:1995ip,Akhmedov:1995vm} realizations, but from the
classification point of view these are only seesaw type-I like diagrams
with additional singlet fields. Many other diagrams are just
propagator corrections, etc. Thus, only very few diagrams lead to {\em
  genuine} neutrino mass models. Here, the word genuine is used for
models which give automatically the leading order contribution to the
neutrino mass at dimension $d$, without the need to introduce additional
symmetries beyond those of the SM.

\begin{figure}[h]
  \begin{center}
  \includegraphics[width=13pc]{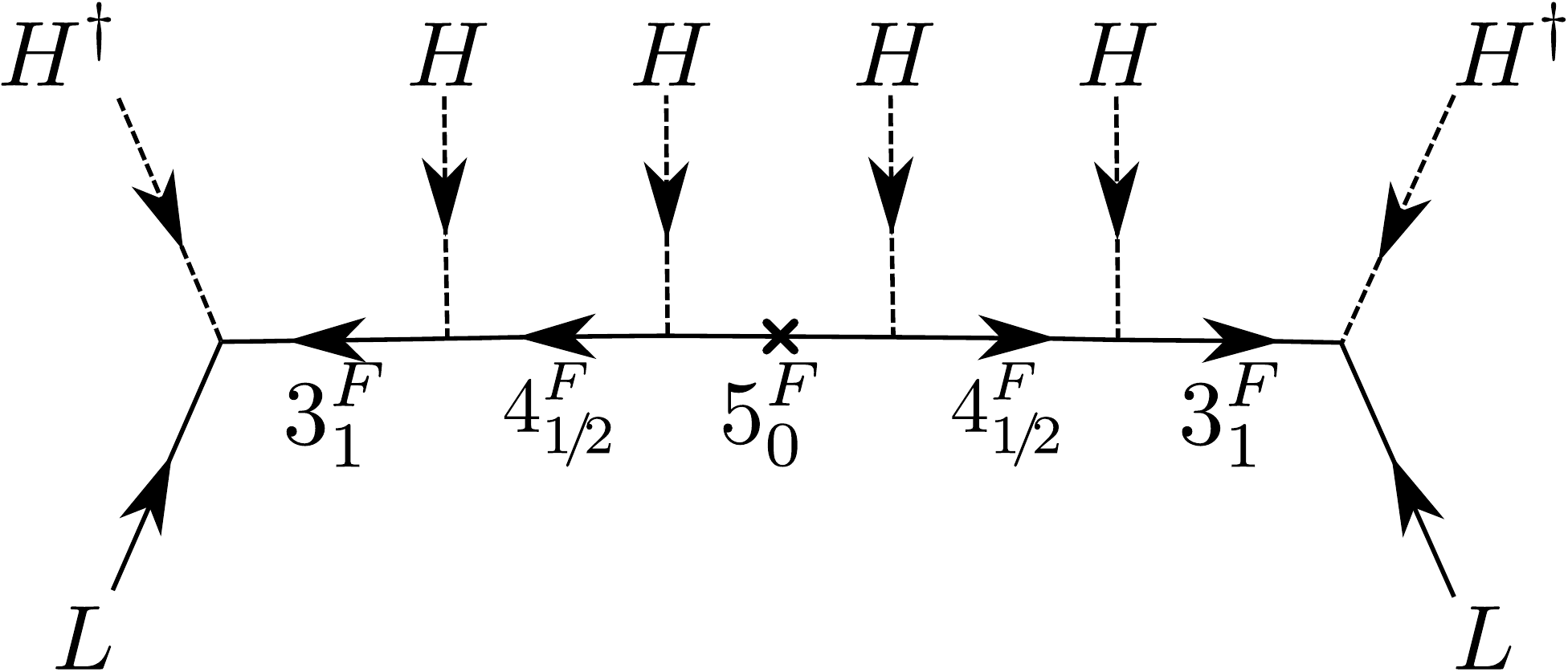}
  \hskip14mm\includegraphics[width=12pc]{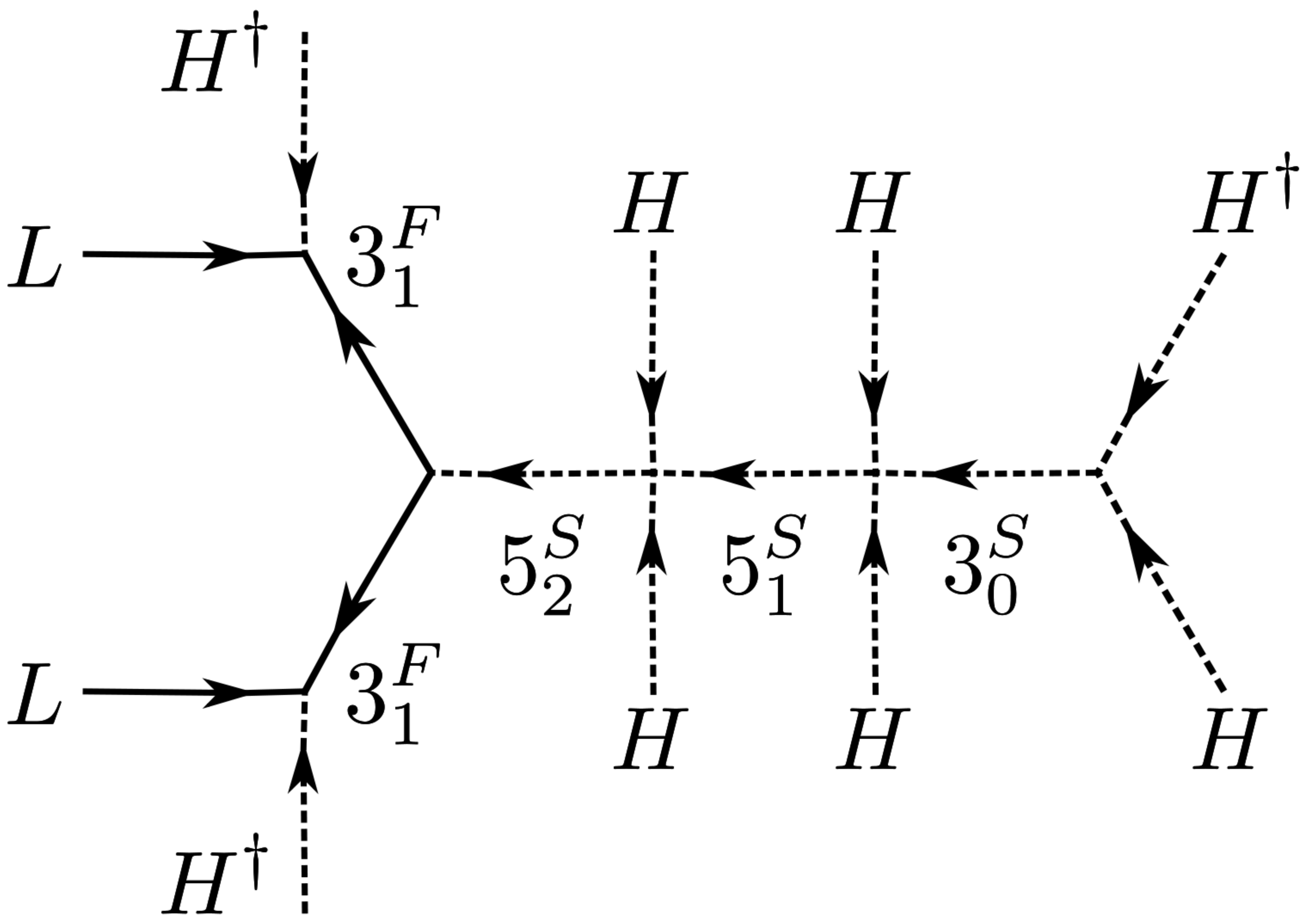}
  \end{center}
    \caption{\label{fig:d9d11}Two examples of genuine models at $d=9$
      (left) and $d=11$ (right).}
\end{figure}

At $d=7$ there is only one genuine model, it was first described in
\cite{Babu:2009aq}. At $d=9$ and $d=11$ there are two possibilities
each; two of them are shown in fig. (\ref{fig:d9d11}).
For the six variations at $d=13$ see \cite{Anamiati:2018cuq}.

\begin{table}
\begin{center}
\lineup
\begin{tabular}{*{6}{l}}
\br                              
Dimension ($d$) & 5 & 7 & 9 & 11 & 13 \cr
Topologies & 1 & 5  & 18 & 92 & 576 \cr
Diagrams & 3 & 9 & 66 & 504 & 4199 \cr
Genuine models & 3 & 1 & 2 & 2 & 6 \cr
\br
\end{tabular}
\end{center}
\caption{\label{tab:topoa}The number of tree-level topologies,
  diagrams and genuine models as function of the dimension $d$ of the
  neutrino mass operator.}
\end{table}

\section{Radiative models and genuineness}

Classical radiative neutrino mass models are the well-known 1-loop Zee
model \cite{Zee:1980ai} and the 2-loop Babu-Zee model
\cite{Cheng:1980qt,Zee:1985id,Babu:1988ki}. At 3-loop, we mention the
KNT \cite{Krauss:2002px} and the AKS models \cite{Aoki:2008av}.

Due to lack of space, here we mention only that the systematic 3-loop
analysis \cite{Cepedello:2018rfh} identified 4367 topologies, from
which however only 44 are genuine and 55 are special genuine, see
below. Five example models are shown in fig. (\ref{fig:exa}). Note
that the 2nd model from the left is the KNT \cite{Krauss:2002px},
which, different from all other models shown here, requires an additional
discrete symmetry in order to avoid a tree-level seesaw contribution to the
neutrino mass.

\begin{figure}[h]
\begin{center}
  \includegraphics[width=7pc]{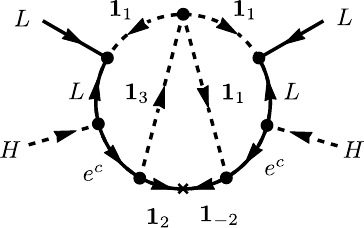}
  \includegraphics[width=7pc]{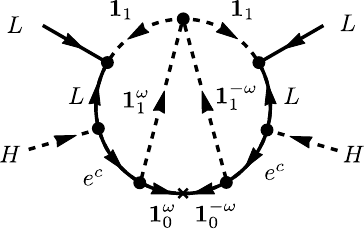}
  \includegraphics[width=7pc]{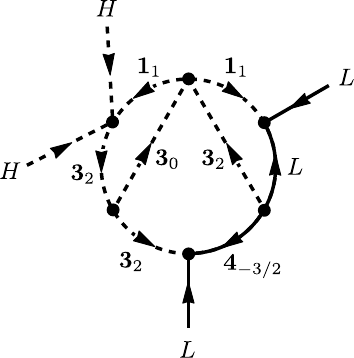}
  \includegraphics[width=7pc]{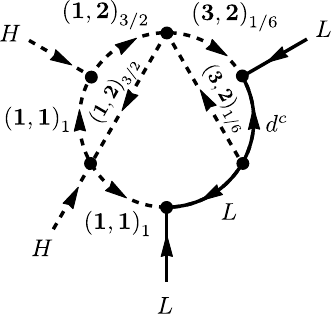}
  \includegraphics[width=7pc]{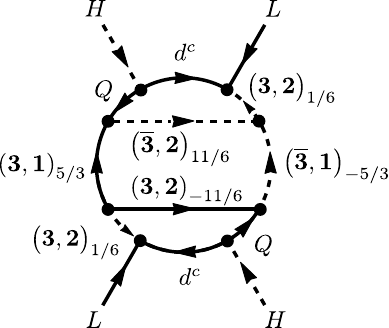}
\end{center}
\caption{\label{fig:exa}Five examples of 3-loop model diagrams.}
\end{figure}

The concept of genuineness, mentioned above, becomes much more subtle
in the case of loop models. This was first discussed in
\cite{Bonnet:2012kz}, where it was shown that some particular diagrams
with finite loop integrals may lead to loop generation of 3-point
vertices, if one adds a $Z_2$ symmetry and a Majorana fermion to the
model in the correct way. In \cite{Sierra:2014rxa} the corresponding
diagrams at 2-loop level were called ``non-genuine, but finite''.
The work of \cite{Cepedello:2018rfh}, however, introduced two more
special cases (class-I and class-II), where diagrams with finite loop
integrals can be associated with genuine models.

\begin{figure}[h]
\begin{center}
  \includegraphics[width=32pc]{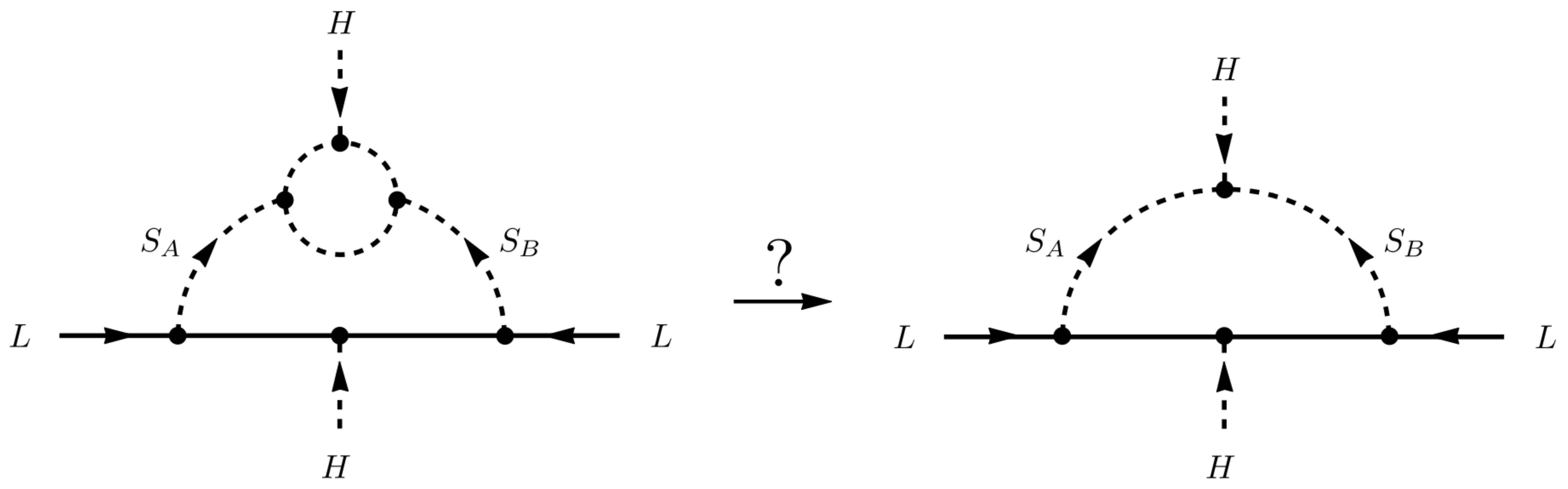}
\end{center}
\caption{\label{fig:spec1}Example of a special genuine diagram,
  class-I.  Here, the loop on the left can be contracted to the vertex
  shown on the right, unless either $S_A$ or $S_B$ is identified with
  the SM Higgs.}
\end{figure}

Two different types of special genuine diagrams have been discussed in
\cite{Cepedello:2018rfh}. While \cite{Cepedello:2018rfh} concerns
itself mostly with 3-loop diagrams, here we will discuss this point
with 2-loop examples. An example of a class-I special genuine diagram
is shown in fig. (\ref{fig:spec1}). Note, that this type of diagram
has been labeled as ISC-i-2 (``internal scalar correction'') in
\cite{Sierra:2014rxa}.  Here, we can see that if the loop connecting
the scalars $S_A$, $S_B$ and the Higgs is allowed by the SM quantum
numbers (or any additional symmetry), then the vertex $S_AS_BH$ on the
right can also not be forbidden by symmetry. Thus, in general one
would conclude that this diagram is non-genuine, although it leads to
a finite loop integral.  However, if either $S_A$ or $S_B$ is
identified with the SM Higgs, the coupling $HHS_B\equiv 0$ vanishes
identically, due to $SU(2)_L$.  In this special case, the 2-loop model
is genuine.

\begin{figure}[h]
  \begin{center}
  \includegraphics[width=12pc]{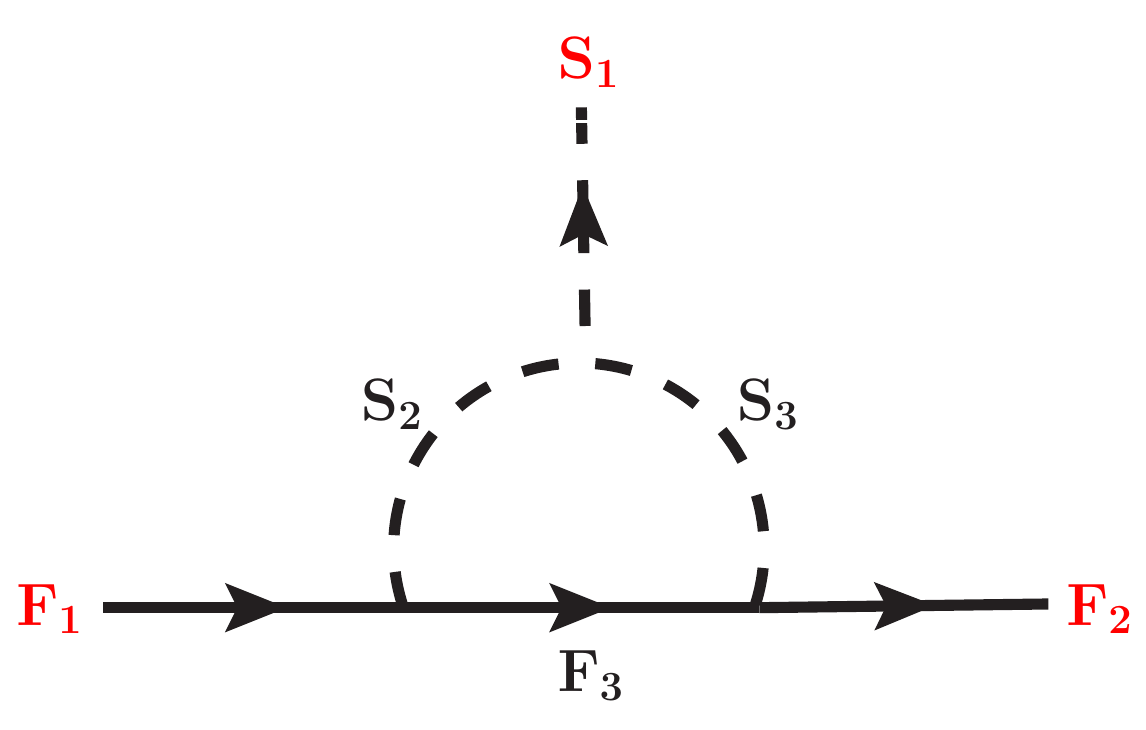}\hskip10mm
  \includegraphics[width=13pc]{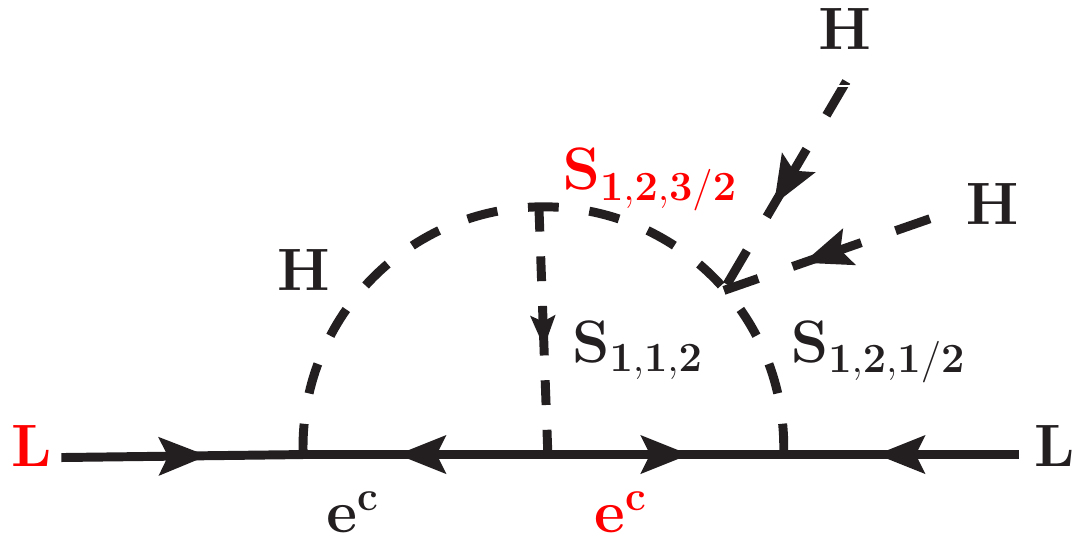}
  \end{center}
  \caption{\label{fig:spec2}To the left: One loop realization of the
    vertex ${\bar F_2}F_1S_1^*$. To the right: Example for a special
    genuine diagram in which the 1-loop generated vertex ${\overline 
      e^c}LS_{1,2,3/2}$ is proportional to a derivative, due to the
    presence of the massless SM fermion in the loop. }
\end{figure}

Class-II of the special genuine diagrams is even more subtle.
Fig. (\ref{fig:spec2}) shows an example. Consider first the 1-loop
diagram on the left. If the internal fermion $F_3$ is massive, its
propagator contains two pieces: $m_F + \slashed{k}$. However, if the
internal fermion in this diagram is massless, then only the term
$\slashed{k}$ exists. In that case, in order to form a mass term, another
momentum term (derivative) can be added at 2-loop level and the
corresponding diagram becomes genuine.  The diagram on the right of
fig. (\ref{fig:spec2}), NG-CLBZ-4 in the notation of
\cite{Sierra:2014rxa}, shows an example.


One can easily show that all 20 ``non-genuine, but finite'' diagrams,
listed in the appendix of \cite{Sierra:2014rxa}, can be identified as
special genuine diagrams in this way.

\section{Summary}

In this paper we have briefly summarized high-dimensional tree-level
neutrino mass models \cite{Anamiati:2018cuq} and the systematic 3-loop
\cite{Cepedello:2018rfh} analysis. We also discussed special
genuine diagrams, not spelled out in detail in \cite{Sierra:2014rxa}. 

\section*{Acknowledgements}

This work was supported by the Spanish grants FPA2017-85216-P and
SEV-2014-0398 (AEI/FEDER, UE), Spanish consolider project MultiDark
FPA2017-90566-REDC, FPU15/03158 (MECD) and PROMETEOII/2018/165 
(Generalitat Valenciana).  R.F. also acknowledges the financial
support from the Grant Agency of the Czech Republic, (GA\v{C}R),
contract nr. 17-04902S, as well as from the grant SEJI/2018/033 (from the Generalitat Valenciana, Spain).

\section*{References}

\providecommand{\newblock}{}

\end{document}